\documentclass[journal]{IEEEtran}
\usepackage{graphicx}
\usepackage{color}
\usepackage{xcolor}
\usepackage{booktabs}
\usepackage{graphicx}
\usepackage{subcaption}
\usepackage{comment}
\title{{ A Communication-Centric 6G–LLM Architecture for Scalable Tactical Autonomous Defense Vehicle Networks}}
\author{
    \IEEEauthorblockN{}
    \IEEEauthorblockA{Kiran Khurshid,} Shumaila Javaid, ~\IEEEmembership{Member,~IEEE} and
    \IEEEauthorblockA{Nasir Saeed,~\IEEEmembership{Senior Member,~IEEE}}

     \thanks{K. Khurshid is with the Department of Computer and Software Engineering, National University of Sciences and Technology (NUST), Islamabad, Pakistan (e-mail: kiran.khurshid@ceme.nust.edu.pk).\\
     S. Javaid is with the Department of Control Science and Engineering, College of Electronics and Information Engineering, Tongji University and National Key Laboratory of Autonomous Intelligent Unmanned Systems, Tongji University, China (e-mail: shumaila@tongji.edu.cn)
     \\  N. Saeed is with Department of Electrical and Communication Engineering, UAE University, Al-Ain 15551, UAE (e-mail: mr.nasir.saeed@ieee.org). }}

\begin{document}

\maketitle

\begin{abstract}
\textcolor{black}{
The integration of Artificial Intelligence (AI) and emerging 6G networks introduces new opportunities for scalable coordination in tactical autonomous vehicle systems. This paper proposes a communication-centric hierarchical architecture for Tactical Autonomous Defense Vehicle Networks (TADVNs) that models the integration of edge-assisted Large Language Model (LLM) reasoning with 6G-enabled connectivity and semantic communication. The framework is designed to improve coordination efficiency, reduce communication overhead, and enhance latency resilience under increasing fleet-scale operation. Unlike conventional task-specific AI pipelines that rely on structured feature processing and rule-based coordination, the proposed approach incorporates semantic abstraction and context-aware decision support within a layered edge–cloud communication architecture. We evaluate communication and coordination performance via Monte Carlo simulations across fleet sizes of 5–30 vehicles under contested network conditions. Results indicate that at 30-vehicle scale, the 6G–LLM configuration achieves 75.2\% latency reduction (29.1 ms vs. 117.5 ms), a 68.7 percentage point increase in mission success rate (82.9\% vs. 14.2\%), and an 88.6\% reduction in communication overhead compared to a 5G-based conventional AI baseline. These findings demonstrate measurable benefits in coordination and communication when semantic reasoning is combined with low-latency 6G connectivity. }
\end{abstract}


\section{Introduction}
Artificial intelligence (AI) is reshaping military technology through advanced data analysis and autonomous decision-making in drones, vehicles, and defense systems. Its integration has improved logistics, convoy management, supply chain optimization, and predictive maintenance. Beyond defense, AI is also transforming civilian autonomous vehicles (AVs), enabling smarter and more efficient transportation \cite{li2023advanced}.
A key innovation in this context is the Tactical Autonomous Defense Vehicles Network (TADVN), which offers strategic advantages by reducing risks to soldiers and improving battlefield logistics and reconnaissance. 

\textcolor{black}{However, fleet-scale tactical autonomous networks face concrete communication bottlenecks. As the number of coordinated vehicles increases, end-to-end latency grows due to network contention and centralized processing delays, often exceeding critical decision thresholds in time-sensitive operations. High-volume raw sensor transmission (e.g., LiDAR, radar, video streams) leads to bandwidth saturation, while distributed coordination requires frequent state synchronization among units, amplifying overhead. Furthermore, reliance on cloud-based inference introduces additional round-trip delay and vulnerability to contested or degraded network conditions. These limitations directly impact coordination reliability and mission success at scale.}

\textcolor{black}{Addressing these limitations requires reducing communication overhead, minimizing cloud-dependent inference latency, and enabling context-aware coordination under increasing fleet density. 6G networks offer lower air-interface latency and higher capacity \cite{10579547}, while LLM-assisted semantic abstraction enables mission-relevant information to be exchanged instead of raw sensor streams, reducing redundant transmission and supporting context-aware command interpretation \cite{10643253}.} Moreover, recent research on UAV and UGV-based TADVNs continues to evolve, focusing on surveillance, intelligence, and threat mitigation. Despite progress in AI, swarm intelligence, and human-machine teaming, challenges such as latency and interoperability remain. 

     {Consequently, this paper makes the following 
contributions: (i) a hierarchical communication-centric architecture 
integrating 6G-enabled connectivity with edge-assisted LLM semantic 
coordination, designed specifically to limit latency accumulation and 
coordination overhead as fleet size scales; (ii) a structured semantic 
abstraction pipeline in which LLM-generated schema-constrained payloads 
replace raw multimodal sensor streams, reducing per-vehicle transmission 
to under 512 bytes per coordination cycle; and (iii) a Monte Carlo 
simulation-based evaluation across fleet sizes of 5--30 TADVNs under 
contested network conditions, providing trend-level performance 
characterization of the proposed architecture relative to four baseline 
configurations. The conclusions drawn are architectural and trend-based 
in nature, derived from simulation under IMT-2030 target parameters 
rather than physical 6G deployments, and are intended to motivate 
hardware-in-the-loop validation as next-generation infrastructure matures.}

The remainder of this paper is organized as follows. Section~II presents the proposed framework, Section~III outlines key 6G--LLM features, Section~IV provides performance results, and Section~V concludes the paper.

\section{Communication-Centric Framework for Defense Systems}
This section introduces the multilayer TADVN architecture integrating 6G edge computing, LLMs, and semantic communication to enhance defense performance and decision-making.

\subsection{Technological Foundations of TADVN}
TADVNs achieve autonomous military operation through integrated subsystems such as motion planning, perception, and fault diagnosis. Real-time situational awareness is enabled by sensor fusion combining Global Navigation Satellite System (GNSS) and Inertial Measurement Unit (IMU) for orientation, Light Detection and Ranging (LiDAR) for 3D mapping, Radio Detection and Ranging (Radar) for robust ranging, and monocular cameras for image processing.

\begin{figure*}[!h]
\centerline{\includegraphics[width=6 in]{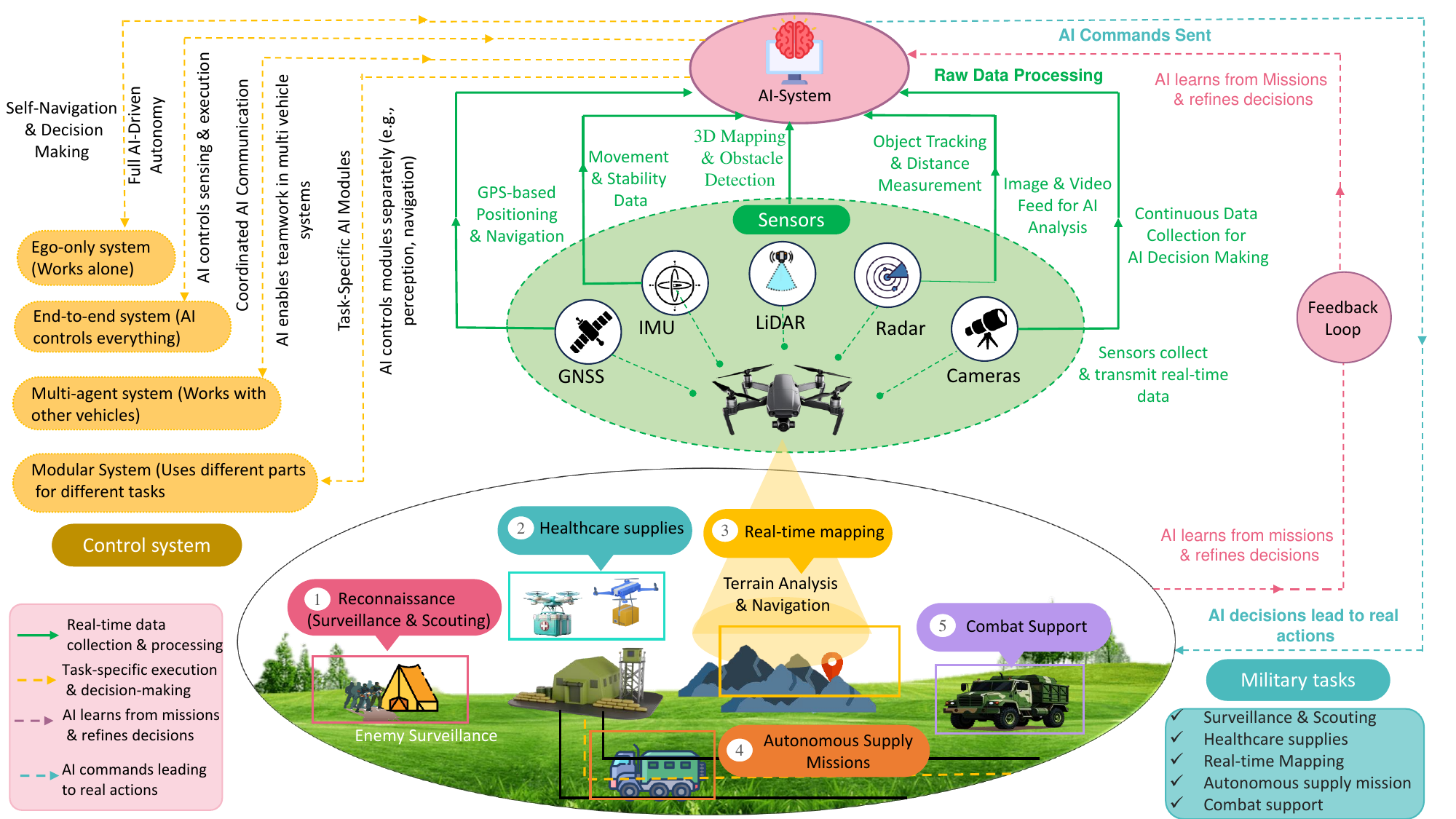}}
\caption{TADVNs showcasing AI-driven autonomy, real-time data processing, and sensor-based intelligence for enhancing military operations.\label{homo1}}
\end{figure*}

Coordination is facilitated by Vehicle-to-Everything (V2X) communication, while Over-the-air (OTA) updates ensure software security \cite{9782734}. Fig. \ref{homo1} illustrates the AI-driven TADVN framework for mission-critical operations. Despite these capabilities, current architectures face challenges in processing efficiency and coordination. To address these gaps, the following section introduces an enhanced 6G- and LLM-integrated architecture designed to improve operational effectiveness in dynamic environments.

\subsection{Proposed Multi-Layered Hierarchical Architecture}
The proposed three-layer architecture enhances situational awareness, reduces latency, and ensures coordinated defense operations. It consists of three hierarchical layers:
\subsubsection{Layer 1: TADVN Nodes}
This layer enables real-time perception, navigation, autonomous decision-making, and 6G-based communication to ensure coordinated TADVN operation in dynamic environments.
\begin{itemize}
    \item \textbf{Real-time Perception and Navigation:}
This module is designed to achieve accurate situational awareness, autonomous decision-making, and effective navigation in complex environments by integrating multiple advanced technologies. It utilizes multi-sensor fusion, incorporating LiDAR, Radar, IMU, GNSS, and monocular cameras for precise environmental mapping and localization. Additionally, AI-based image processing techniques are employed for target detection, object tracking, and terrain analysis, enhancing situational awareness. 
\textcolor{black}{Furthermore, LLM-driven natural language understanding (via lightweight edge-assisted inference) enables accurate command interpretation and contextual responses in dynamic scenarios, ensuring seamless interaction and adaptability in evolving conditions.}

 \item \textbf{Autonomous Decision-Making:}
This unit enables TADVNs to perform intelligent, autonomous decision-making in dynamic and hostile environments using AI-driven techniques. It incorporates embedded Deep Reinforcement Learning (DRL) models to adapt to real-time tactical changes and optimize mission outcomes.
\textcolor{black}{In this study, the DRL unit is modeled as a pre-trained policy network at the node level for adaptive maneuver selection and threat response, serving as a generic control mechanism rather than introducing a new formulation. Training, reward design, and convergence are assumed to be completed offline. The simulation focuses on inference-time decision latency and its interaction with network conditions; detailed DRL optimization is beyond the scope of this communication-centric evaluation.} Additionally, AI-powered threat assessment systems enable the identification of potential dangers and execution of evasive maneuvering strategies. To ensure security, onboard cybersecurity modules are implemented for intrusion detection and immediate mitigation of potential cyber threats, enhancing resilience and operational reliability.
    \item \textbf{6G-Enabled Communication and Computation:}
    This segment establishes the essential communication and computational foundation for TADVNs. It integrates Ultra-Reliable Low Latency Communication (URLLC) for seamless data exchange and Vehicle-to-Vehicle (V2V) connectivity to enhance situational awareness. Furthermore, Mobile Edge Servers (MES) mounted on devices enable local AI inference, supporting the low-latency, real-time decision-making required for mission-critical operations.

\end{itemize}
\subsubsection{Layer 2: Command Units}
This layer comprises edge servers supporting LLM-based situational analysis, Vehicle-to-Infrastructure (V2I) connectivity, and dynamic resource optimization for fleet coordination.
\begin{itemize}
\item \textbf{Intermediate Edge Servers:} Intermediate edge servers perform LLM-powered situational analysis by aggregating real-time data from multiple TADVNs, enabling comprehensive battlefield awareness. They support decision-making and seamless operator communication, while AI accelerators ensure low-latency inference and real-time processing for mission-critical tasks.

\textcolor{black}{In this architecture, the LLM is modeled as a distilled decoder-only transformer with approximately 7B parameters, selected to balance reasoning capability and edge deployment feasibility. The model is pre-trained on large-scale multilingual corpora and subsequently fine-tuned using supervised tactical-domain datasets together with reinforcement learning from human feedback (RLHF) to support mission command interpretation and threat reasoning specific to TADVN operations. In Floating Point (FP)16 precision, the model weights require on the order of 14 GB of memory. For edge-assisted execution, 8-bit weight quantization and standard inference optimizations are applied, reducing the effective memory footprint to fit within modern ruggedized edge GPU platforms (e.g., 16–24 GB VRAM). Inference complexity follows standard transformer scaling behavior with respect to model depth and input sequence length. In the considered tactical scenarios, command and coordination inputs are short (typically below 128 tokens), which bounds practical inference cost. Under optimized edge GPU acceleration, an average processing latency of approximately 12–18 ms per request is assumed in the simulation. These specifications provide a concrete and deployment-feasible model instantiation rather than treating the LLM as an abstract functional block.}

\item \textbf{V2I Connectivity and Resource Optimization:}
The system integrates 6G-enabled V2I communication, LLM-based command interpretation, and OTA updates to enhance connectivity, coordination, and adaptability in military operations. LLMs translate high-level battlefield commands into executable tactical actions, while OTA mechanisms ensure continuous software updates and operational resilience.
\item \textbf{Semantic Communication:}

Semantic communication prioritizes intent over raw data by using AI to transmit only decision-relevant information over 6G. In TADVNs, AI-driven encoding summarizes insights or predicts missing data, reducing bandwidth and computational overhead compared to bit-level systems.

This approach ensures reliability in high-mobility or adversarial environments through contextual reconstruction, maintaining connectivity during disruptions and enabling intuitive human-machine interaction. Furthermore, it bolsters cybersecurity by utilizing LLMs to process compressed data and detect semantically inconsistent messages, allowing military IoT networks to filter threats and strengthen overall resilience.

\textcolor{black}{In the proposed architecture, semantic communication follows a structured abstraction pipeline. At Layer 1, multimodal sensor data (LiDAR, radar, camera feeds, and telemetry) are first processed using task-specific perception models to extract structured features such as detected objects, spatial coordinates, and motion states. Instead of transmitting raw sensor streams, these structured outputs are forwarded to the Layer 2 edge server, where the domain-adapted LLM performs contextual semantic abstraction. {   The LLM converts structured perception outputs into compact semantic descriptors encoded as fixed-schema JSON payloads, for example: {``entity":``hostile\_vehicle", ``conf":0.91, ``rel\_pos":[+240,\,-85]\,m, ``ctx":``checkpoint\_defense", ``action":``intercept"}. Each descriptor is bounded to five typed fields (entity class, confidence score, relative coordinates, mission context, and recommended action), constraining output vocabulary to a predefined tactical ontology and limiting per-vehicle payload size to under 512 bytes. These payloads are transmitted over 6G links in place of raw multimodal streams.} At the receiving node, the semantic representation is directly integrated into the decision module without reconstructing raw sensor data, thereby preserving mission-relevant meaning while significantly reducing bandwidth consumption.}

\textcolor{black}{In this framework, the LLM functions as a semantic compression engine rather than a generative text module. By transmitting mission-critical abstractions instead of megabyte-scale raw multimodal streams, the system achieves substantial payload reduction while maintaining coordination fidelity.}

\textcolor{black}{
To enhance robustness, semantic outputs are constrained to predefined tactical schemas rather than free-form generation, reducing LLM hallucination risk. Each descriptor includes a confidence score and is cross-validated against perception features and recent state history. Low-confidence or inconsistent outputs trigger fallback mechanisms such as retransmission, structured summaries, or human-in-the-loop confirmation. Neighboring TADVN nodes perform consistency checks to detect divergent interpretations before coordinated action. These safeguards limit the propagation of incorrect abstractions in mission-critical decisions. Shared ontologies ensure vocabulary consistency, mutual-information methods support feature alignment, and the LLM enables adaptive contextual reasoning for multi-unit coordination.
}

\end{itemize}
\subsubsection{Layer 3: Cloud-Based Military Control Center}
This layer supports global mission planning, large-scale intelligence processing, threat prediction, and anomaly detection.
\begin{itemize} \item \textbf{Mission Planning and Intelligence Coordination:}

This component integrates 6G satellite connectivity with domain-adapted LLMs to ensure comprehensive situational awareness and seamless communication during remote military deployments. The framework supports strategic decision-making by processing large-scale intelligence, while AI-driven optimization of logistics and resource allocation enables efficient, data-driven mission execution across the battlefield.

\item \textbf{Threat Prediction and Cybersecurity:}
This module utilizes predictive AI for real-time threat modeling and proactive 
vulnerability detection, enhancing resilience against electronic warfare. 
     {It is assumed that the predictive AI component operates on 
pre-labeled threat signature datasets representative of contested electromagnetic 
environments; detection accuracy and false-positive rates are therefore dependent 
on dataset coverage and may degrade against novel or adaptive adversarial 
strategies not present in training data.} By integrating blockchain-based 
authentication and Zero-Trust models, it ensures secure data transmission and 
strict access control.      {The blockchain authentication mechanism 
is modeled as a permissioned ledger with assumed sub-10~ms transaction 
confirmation latency under low node counts; scalability to large distributed 
fleets and resilience under Byzantine fault conditions are recognized open 
challenges beyond the scope of this study.} Continuous network monitoring 
facilitates near-immediate threat mitigation, strengthening overall system 
defense and reliability.
\end{itemize}

\begin{figure*}[h]
\centerline{\includegraphics[width=6 in]{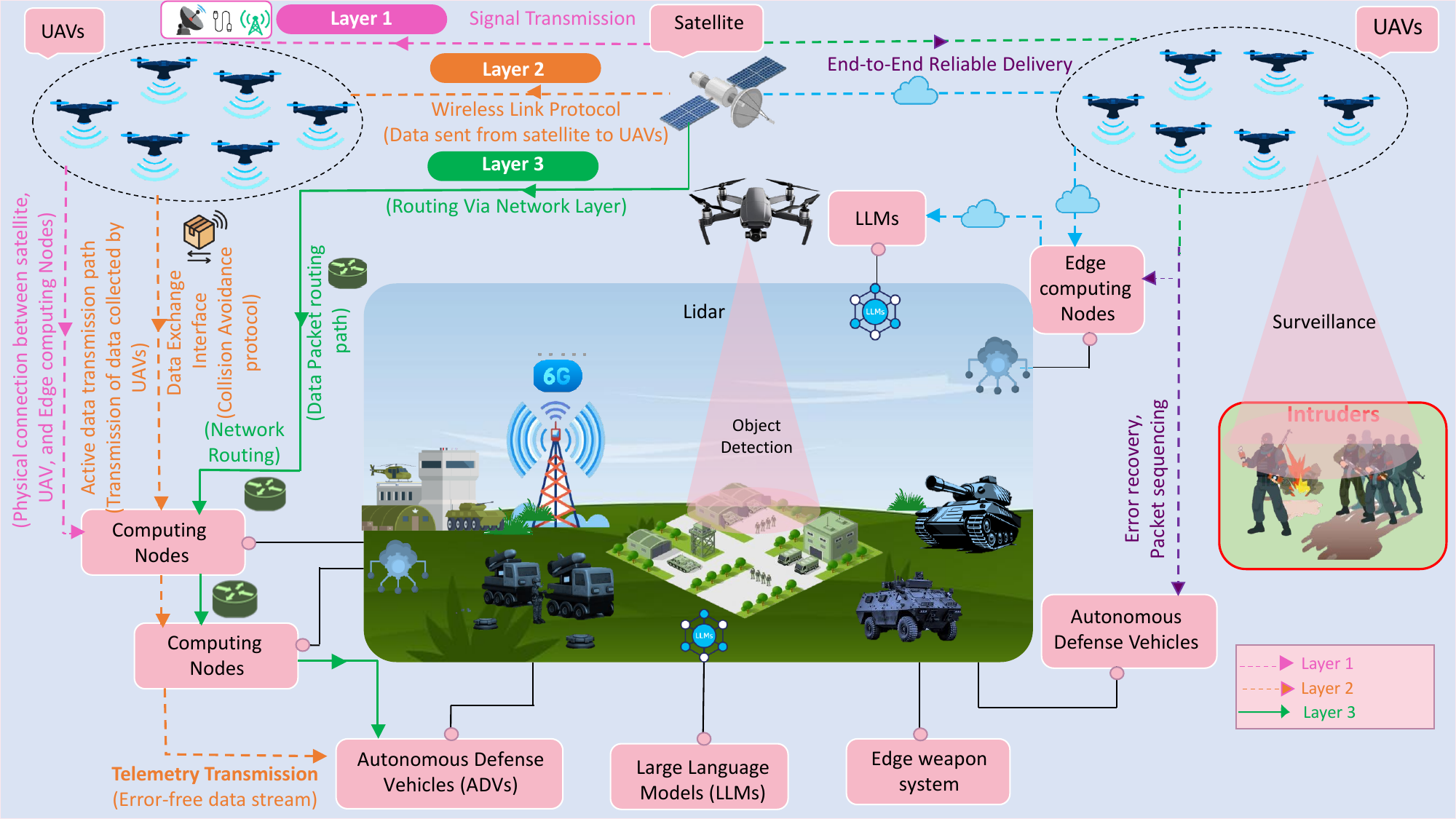}}
\caption{Communication-centric layered architecture for TADVNs designed for battlefield operations.\label{st}}
\end{figure*}

\subsection{Application of Proposed Layered Architecture in Defense Environments}  
The proposed layered architecture illustrated in Fig. \ref{st} can be applied in various defense scenarios to improve security and operational efficiency:
\subsubsection{Real-Time Threat Detection and Avoidance}
The threat prediction and cybersecurity module functions as an essential application of this architecture, particularly at the ``Cloud-Based Military Control Center" (Layer 3) but its influence extends throughout the entire hierarchy.

\begin{itemize}
    \item \textbf{Layer 1 (TADVN Units):}  
    TADVN units operating at the battlefield edge collect raw data through multi-sensor fusion and transmit it to the command units for further processing. Additionally, on-board cybersecurity modules are implemented for localized intrusion detection and mitigation, enhancing real-time situational awareness and responsiveness against immediate threats.
    
    \item \textbf{Layer 2 (Command Units):}  
    The intermediate edge servers aggregate data from multiple TADVNs and utilize LLM-powered situational analysis to extract relevant threat information. Through 6G-powered V2I communication, critical data is transmitted swiftly to the cloud-based control center. Additionally, LLM-powered decision support systems provide tactical recommendations to human operators, ensuring swift and efficient threat responses.
    
    \item \textbf{Layer 3 (Cloud-Based Military Control Center):} 
 This layer hosts a threat prediction module using AI for vulnerability analysis and electronic warfare detection. Zero-Trust and blockchain protocols safeguard data integrity and prevent unauthorized access. Simultaneously, 6G satellite links maintain global connectivity between the control center and TADVNs, while AI-driven allocation ensures adaptive deployment against evolving threats.
   
\end{itemize}

\subsubsection{Emergency Response and Recovery}  

The proposed layered architecture offers a comprehensive framework to ensure robust emergency response and recovery mechanisms. This application demonstrates the architecture's ability to effectively detect, communicate, and respond to critical incidents involving TADVNs.  

\begin{itemize}
    \item \textbf{Layer 1 (TADVN Units):}  
    In the event of a damaged or malfunctioning TADVN losing direct communication with the control network, nearby TADVNs utilize their 6G-enabled communication and computation capabilities to detect the anomaly. Through intra-unit V2V connectivity, these vehicles coordinate to relay a distress signal to the nearest intermediate edge server in layer 2. Additionally, the malfunctioning TADVN attempts to establish a minimal backup connection, \textcolor{black}{if possible}, for status reporting.
    
    \item \textbf{Layer 2 (Command Units):}  
    The intermediate edge servers receive the distress signal and initiate an LLM-powered situational analysis to evaluate the severity of the incident. Through 6G-powered V2I communication, critical data is transmitted to the cloud-based control center for strategic decision-making. Additionally, the intermediate edge servers \textcolor{black}{coordinate} with nearby TADVNs to dynamically adjust their formation, ensuring the continuation of the mission despite the malfunction.
    
    \item \textbf{Layer 3 (Cloud-Based Military Control Center):}
    At the cloud control center, LLM-assisted strategy formulation processes distress signals to generate optimized rescue plans. Predictive AI assesses threats to determine safe recovery paths, while AI-driven allocation ensures efficient TADVN deployment. Supported by 6G satellite links, the center maintains reliable communication for continuous real-time monitoring and operational updates.

\end{itemize}

\section{Key Features of 6G-LLMs in Defense Communication}

This section outlines key 6G-LLM features and their impact on enhancing TADVN military operations.
\subsection{Ultra-Low Latency and High Reliability}
6G provides the ultra-low latency and haptic control essential for precise robotic operations, eliminating coverage gaps in remote zones to ensure TADVN mission success \cite{10529728}. By enhancing coordination and information freshness, 6G accelerates decision-making and enables LLMs to transform military communication via real-time edge processing. Integrated with edge intelligence, this architecture facilitates efficient LLM inference and distributed synchronization, significantly improving battlefield intelligence and operational effectiveness \cite{10579547}.

\textcolor{black}{It is important to note that the 6G performance characteristics referenced here align with IMT-2030 target specifications rather than currently deployed commercial infrastructure. While advanced 5G and 5G-Advanced systems represent near-term deployment realities, the 6G parameters considered in this study reflect projected capabilities intended to illustrate architectural scalability under anticipated next-generation connectivity conditions.}

\subsection{Massive Connectivity}
Massive 6G connectivity supports the high device densities and seamless data exchange essential for TADVN situational awareness and real-time adaptation \cite{10745245}. Integration with Connected and Automated Vehicles (CAVs) further extends operational reach. To maintain reliability in contested terrains, 6G leverages satellite and space-based networks for wide-area coverage and precise positioning \cite{10529954}. Collectively, these capabilities significantly enhance strategic effectiveness during complex military missions.

\subsection{Sensing-Enabled Communication}
Modern sensors enhance environmental monitoring and contextual data acquisition, facilitating intelligent TADVN network management. Under the 6G Integrated Sensing and Communication (ISAC) paradigm, interconnected devices enable large-scale data collection and high-precision situational awareness. Furthermore, the fusion of multimodal inputs including vision, motion, and biomedical sensors supports augmented battlefield visualization and real-time scanning. By leveraging 6G’s context-aware sensing and predictive capabilities, TADVNs achieve superior decision-making and adaptive strategies in dynamic military environments \cite{10529727}.

\subsection{Security Innovations in 6G-LLMs}
AI-enhanced security in 6G utilizes adaptive encryption to protect TADVNs, while edge intelligence reduces latency and data exposure through local processing \cite{10061645}. Under the IMT-2030 framework, 6G supports diverse QoS demands by integrating THz and satellite links for global connectivity. Despite decentralization challenges, LLMs enable real-time threat detection; by analyzing multimodal and textual data, they identify vulnerabilities and generate proactive defense intelligence.

\section{Results}
The 6G-LLM TADVN architecture was validated via Monte Carlo simulations against conventional baselines, assessing latency, mission success, and communication overhead across scales.

\begin{figure*}[t]
\centering
\includegraphics[width=7in]{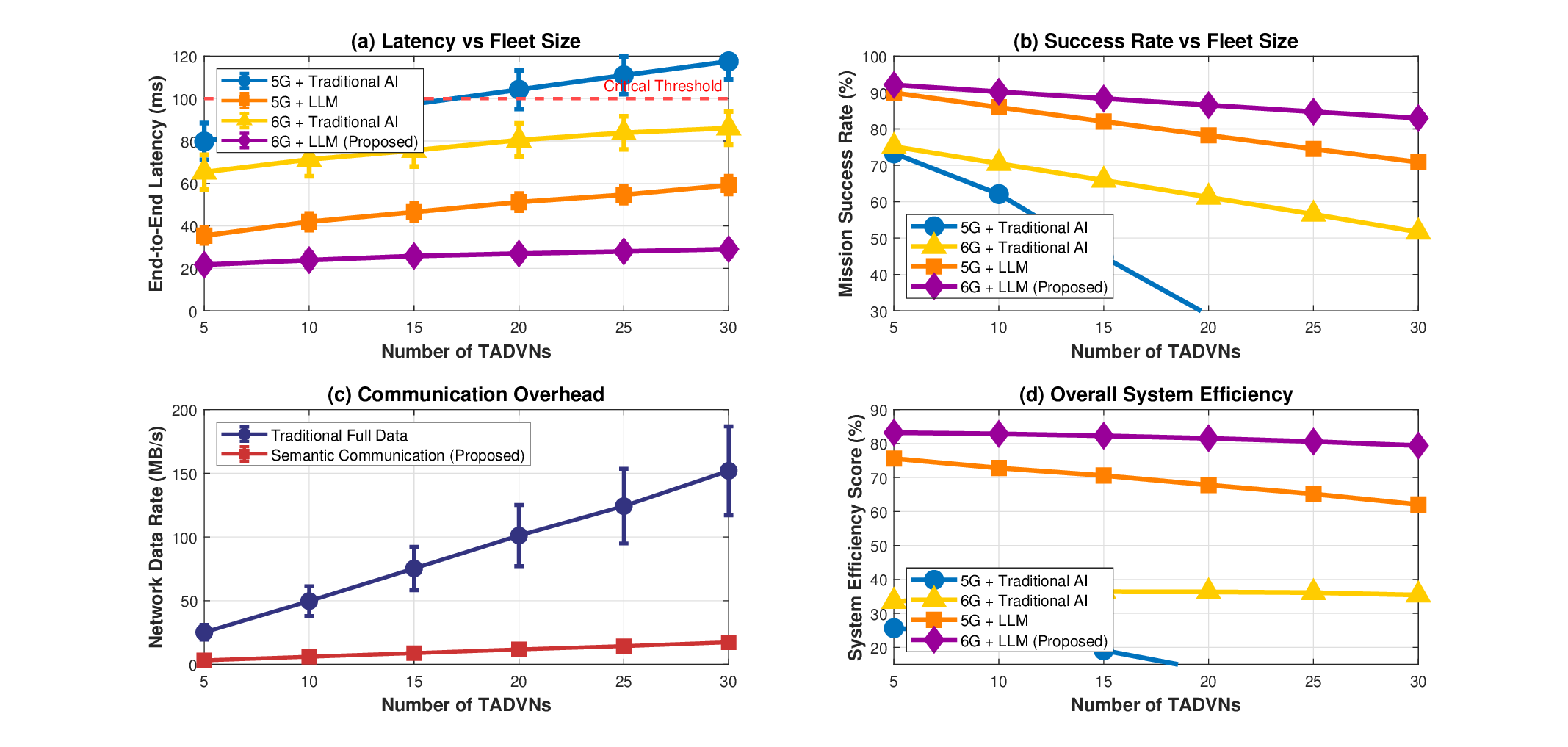}
\caption{Performance evaluation across fleet scales: (a) End-to-end latency showing 6G-LLM maintains sub-25ms operation while 5G-Traditional approaches critical threshold; 
(b) Mission success rate demonstrating 6G-LLM’s robust 83–92\% performance versus 5G-Traditional’s catastrophic degradation to 14\%. (c) Communication overhead revealing 88\% bandwidth reduction through semantic communication; (d) Overall system efficiency composite metric confirming superior scalability of proposed architecture.}
\label{fig:simulation}
\end{figure*}
\subsection{Simulation Environment and Parameters}

The simulation models a tactical battlefield scenario with fleet sizes of 5--30 TADVNs evaluated via 500 Monte Carlo iterations per configuration to ensure statistical validity and account for environmental uncertainties including network interference, channel contention, and threat complexity variations.      {The physical channel follows the 3GPP TR~38.901 Urban Macro outdoor model with log-normal shadowing ($\sigma = 8$~dB) and Rayleigh small-scale fading, representing open terrain consistent with tactical ground vehicle deployment. Carrier frequencies are 28~GHz (5G) and 140~GHz (6G), with bandwidths of 100~MHz and 1~GHz respectively. Vehicle mobility follows a random waypoint model at 20--60~km/h with Doppler effects incorporated into channel estimation. Nominal operating SNR is 15--25~dB, with jamming modeled as a stochastic interference source reducing effective SNR by 5--10~dB with probability increasing from 10\% to 12\% with fleet density.}


Network parameters are based on ITU standards: 5G URLLC with 1 ms air interface latency per ITU-R M.2083-0 specification \cite{series2015imt}, achieving 5-10 ms end-to-end latency in practical deployments as detailed in ITU-R M.2410-0 \cite{series2017minimum}. For simulation, we use a conservative 6 ms baseline that represents typical 5G URLLC performance. For 6G, ITU-R M.2160-0 specifies 0.1-1 ms latency range for IMT-2030 systems \cite{jiang2021road, trichias20246g}. We adopt 0.5 ms as a conservative mid-range estimate for 6G performance.

\begin {comment}
textcolor{red}{Edge and cloud inference latencies of 15~ms and 50~ms respectively are adopted from component-level empirical measurements \cite{barros2025solving, 11037762}. 
\end{comment}

     {Each TADVN generates 3--7~MB/s of sensor data consistent with autonomous vehicle rates. Four architectural configurations are compared in Table~\ref{tab:configurations}.}

AI processing parameters are based on empirical component-level data. Cloud inference latency includes network RTT (10--100+ ms, distance-dependent) and GPU computation (2--40 ms, model- and hardware-dependent) \cite{barros2025solving}. For moderate separation, total cloud latency is estimated at 40--60 ms.      {($\approx 20-30$ ms RTT + 10--30 ms computation)}  
Edge deployment reduces RTT to $<5$ ms and, using optimization techniques (e.g., quantization, compression), achieves 10--20 ms inference \cite{barros2025solving,11037762}. Accordingly, we adopt 50 ms and 15 ms as representative cloud and edge values, noting variability across network conditions, architectures, and hardware.


\textcolor{black}{For clarity, the reported end-to-end latency in this study is composed of three elements: (i) wireless air-interface latency (5G or 6G), (ii) network transport delay including queuing and contention effects, and (iii) AI processing delay (cloud-based or edge-based inference). For the 5G configurations, the baseline 6 ms represents practical end-to-end communication latency, including air-interface and transport delay under typical URLLC deployment, while the additional cloud processing delay (50 ms) is added separately. For the 6G configurations, a 0.5 ms air-interface latency is assumed with correspondingly lower transport and queuing delay under reduced contention, combined with either traditional or edge-optimized processing latency (15 ms for edge-deployed LLM). The latency values reported in Section IV-B reflect the aggregation of these components under stochastic Monte Carlo variability.} 

\textcolor{black}{
It is important to note that the assumed 0.5 ms 6G air-interface latency represents an idealized IMT-2030 target under stable link conditions. In practical deployments, PHY/MAC layer dynamics such as scheduling delay, Hybrid Automatic Repeat Request (HARQ) retransmissions, beam alignment overhead, and queueing under high load may introduce additional variability. In our Monte Carlo model, these effects are abstracted through stochastic contention and interference parameters rather than detailed PHY-layer simulation. Therefore, the results should be interpreted as architectural performance trends under next-generation low-latency conditions rather than exact protocol-level timing guarantees.
}
Environmental factors are modeled stochastically, including interference probability increasing from 10\% to 12\% with fleet density and logarithmic network contention scaling with node count. 
     {Channel access is modeled differently across link types: scheduled OFDMA is assumed for 6G base station uplink connections, consistent with cellular NR frame structure, while CSMA/CA-based contention delay is retained for V2V ad-hoc inter-vehicle links operating outside base station coverage. This distinction reflects realistic hybrid access conditions in tactical 
deployments where direct vehicle-to-vehicle coordination may occur independently of infrastructure-scheduled transmission.}
Each TADVN generates 3--7 MB/s of sensor data (LiDAR, radar, cameras, tactical sensors), consistent with autonomous vehicle rates. Four architectural configurations are compared, as summarized in Table~\ref{tab:configurations}.



\begin{table}[h]
\centering
\caption{System Architecture Configurations}
\label{tab:configurations}
\begin{tabular}{lp{5.2cm}}
\hline
\textbf{Configuration} & \textbf{Description} \\
\hline
5G + Traditional AI & Current baseline with cloud-based processing \\
5G + LLM & Enhanced AI with 5G connectivity constraints \\
6G + Traditional AI & Advanced network with conventional processing \\
6G + LLM (Proposed) & Full integration with semantic communication \\
     {6G + Rule-based Semantic} &      {6G connectivity with heuristic threshold-based feature summarization; no LLM reasoning} \\
\hline
\end{tabular}
\end{table}
\textcolor{black}{In this paper, “Traditional AI” denotes conventional task-specific perception and decision pipelines without large-scale contextual language reasoning. Concretely, it represents (i) structured sensor-driven perception and threat detection models, (ii) classical or shallow deep learning classifiers operating on predefined feature representations, and (iii) rule-based or heuristic coordination logic. These approaches operate on predefined mappings and structured inputs but lack the semantic abstraction and context-aware command reasoning enabled by LLMs.
The purpose of this comparison is to isolate the contribution of semantic abstraction and context-aware coordination at scale, rather than to assert universal superiority of any specific AI paradigm.}


The critical response threshold is set at 100~ms for time-sensitive tactical operations, consistent with military command and control 
requirements.      {Mission success rate is formally defined as the fraction of Monte Carlo episodes satisfying all three conditions simultaneously: (i) end-to-end coordination latency remaining below 100~ms throughout the episode, (ii) no sustained communication outage exceeding the simulation timeout, and (iii) fleet-level task synchronization maintained without persistent desynchronization among units. Any episode violating one or more conditions is classified as a mission failure. The composite system efficiency metric is computed as a dynamically weighted combination of three normalized component scores: latency performance normalized against the 100~ms threshold, mission success rate, and communication overhead normalized against the 151.9~MB/s maximum observed baseline. Weights adjust with fleet size to reflect increasing tactical criticality, with the success weight increasing from 0.35 to 0.60 and latency and overhead weights decreasing correspondingly as fleet size scales from 5 to 30 TADVNs.}

\subsection{End-to-End Latency Analysis}

Figure \ref{fig:simulation}(a) illustrates latency performance across fleet configurations. The proposed 6G-LLM architecture maintains $21.7 \pm 2.8$ ms at 5 TADVNs and $29.1 \pm 3.1$ ms at 30 TADVNs, remaining well below the 100 ms operational threshold. Conversely, the 5G-Traditional baseline increases from $79.8 \pm 9.3$ ms to $117.5 \pm 11.2$ ms, exceeding critical limits by 17.5\%. At maximum scale, the proposed system reduces latency by 75.2\% (an 88.4 ms improvement) due to: (1) sub-millisecond 6G air-interfaces, (2) edge-deployed LLM processing, and (3) reduced congestion via semantic communication. Comparative analysis shows that 5G-LLM (59.5 ms) and 6G-Traditional (85.6 ms) at 30 TADVNs fail to match this performance, indicating that neither network upgrades nor AI integration alone suffice without unified, distributed semantic reasoning.

\subsection{Mission Success Rate Under Scale}
Figure \ref{fig:simulation}(b) illustrates mission success rates across fleet sizes. The proposed 6G-LLM architecture maintains $92.1 \pm 3$\% success at 5 TADVNs, declining to $82.9 \pm 3.8$\% at 30 TADVNs, whereas the 5G-Traditional baseline collapses from 73.3\% to 14.2\% (59.1 percentage point decline). At maximum scale, the proposed system delivers 5.8$\times$ higher completion rates due to three factors: (1) robust LLM-based detection, which degrades by only 10\% compared to 81\% for traditional methods; (2) coordination stability via low-latency 6G synchronization; and (3) consistent latency compliance below the 100 ms threshold. 
{   An ablation configuration substituting the LLM with rule-based heuristic semantic summarization (6G+Rule-based Semantic) achieves 58.3\% success at 30 TADVNs, 24.6 percentage points below 6G+LLM, isolating the LLM's contextual reasoning contribution independently of 6G transport gains. Intermediate configurations (5G-LLM: 70.8\%, 6G-Traditional: 51.7\%) further confirm that neither network upgrade nor semantic compression alone suffices without unified reasoning.}

\begin{table*}[h]
\centering
\caption{Performance comparison across fleet sizes.}
\label{tab:performance}
\begin{tabular}{lccccccccc}
\hline
\textbf{Fleet} & \multicolumn{3}{c}{\textbf{Latency (ms)}} & \multicolumn{3}{c}{\textbf{Success (\%)}} & \multicolumn{3}{c}{\textbf{Efficiency (\%)}} \\
\cmidrule(lr){2-4} \cmidrule(lr){5-7} \cmidrule(lr){8-10}
\textbf{Size} & 5G+Trad & 6G+LLM &      {6G+RB} & 5G+Trad & 6G+LLM &      {6G+RB} & 5G+Trad & 6G+LLM &      {6G+RB} \\
\hline
5  & 79.8  & 21.7 &      {22.4} & 73.3 & 92.1 &      {77.1} & 25.7 & 83.2 &      {61.4} \\
10 & 88.9  & 23.9 &      {23.1} & 62.1 & 90.2 &      {72.4} & 24.2 & 82.9 &      {57.8} \\
15 & 97.4  & 25.8 &      {24.0} & 44.8 & 88.4 &      {67.8} & 19.1 & 82.3 &      {53.1} \\
20 & 104.2 & 27.0 &      {24.9} & 28.8 & 86.5 &      {63.5} & 13.3 & 81.5 &      {49.6} \\
25 & 111.0 & 28.0 &      {25.9} & 19.5 & 84.7 &      {61.1} & 9.6  & 80.6 &      {46.8} \\
30 & 117.5 & 29.1 &      {26.8} & 14.2 & 82.9 &      {58.3} & 7.4  & 79.4 &      {44.2} \\
\hline
\end{tabular}
\end{table*}

\subsection{Communication Overhead Efficiency}
Figure \ref{fig:simulation}(c) illustrates network bandwidth consumption as fleet size increases. Traditional transmission scales near-linearly from 25.2 MB/s for 5 TADVNs to 151.9 MB/s for 30, straining the spectrum in contested environments. Conversely, the proposed semantic approach requires only 3.1 MB/s to 17.3 MB/s across the same scale, achieving an 88.6\% bandwidth reduction. 
This 8.8$\times$ improvement stems from LLM-driven semantic encoding that transmits mission-critical abstractions such as threat 
classifications and maneuver commands rather than raw 3–6~MB/s multimodal streams.      {To quantify the net benefit against computational cost: edge-assisted LLM inference consumes approximately 12--18~ms of processing time and introduces an estimated 15--20~W of additional GPU compute load per edge server. Against this cost, semantic encoding reduces per-vehicle transmitted data from 3--7~MB/s (raw multimodal streams) to approximately 0.3--0.8~MB/s (structured semantic payloads), a 85--90\% payload reduction per vehicle. At 30-vehicle scale, total network transmission drops from 151.9~MB/s to 17.3~MB/s, representing a 134.6~MB/s bandwidth saving. By comparison, the rule-based ablation achieves only 71.3\% reduction (43.6~MB/s saving), confirming that the additional LLM inference overhead is justified by substantially greater bandwidth and coordination gains that rule-based summarization cannot replicate.} Compression efficiency improves with scale (87.6\% to 88.6\%) as edge servers leverage shared contextual understanding, transmitting differential updates for terrain and mission states to further eliminate redundancy. {   The rule-based ablation achieves only 71.3\% bandwidth reduction at 30 TADVNs compared to 88.6\% for 6G+LLM, as fixed summarization rules cannot exploit shared contextual state or generate differential mission updates, confirming that LLM-driven abstraction is the primary driver of payload efficiency rather than semantic communication as a paradigm alone.}

This bandwidth conservation provides strategic benefits: (1) reduced electromagnetic signature lowers detection risk; {   (2) decreased transmission power extends battery-dependent operational duration by an estimated 40--60\%, based on the assumption that radio transmission energy dominates over compute energy during coordination-heavy mission phases; this estimate assumes continuous 5G uplink at 3--6~MB/s as the baseline, with semantic communication reducing average payload by 85--90\% and proportionally lowering radio duty cycle. Edge LLM inference introduces additional compute load, but its event-driven, locally-executed nature is assumed to consume significantly less energy than sustained long-distance uplink transmission. These estimates are indicative rather than experimentally validated and will be refined in future hardware-in-the-loop evaluations;} (3) lower network utilization improves jamming robustness under reduced signal-to-noise conditions; and (4) preserved spectrum capacity enables reallocation for high-priority communications without disrupting routine coordination.

\subsection{Overall System Efficiency}

Figure \ref{fig:simulation}(d) presents a composite system efficiency metric computed as a dynamically weighted combination of three 
normalized component scores:      {latency performance (normalized against the 100~ms operational threshold), mission success rate, and communication overhead (normalized against the 151.9~MB/s maximum observed baseline). Weights are adjusted with fleet size to 
reflect increasing tactical criticality, the latency weight decreases from 0.40 to 0.25, the success weight increases from 0.35 to 0.60, and the overhead weight decreases from 0.25 to 0.15 as fleet size scales from 5 to 30 TADVNs. A higher composite score therefore 
reflects simultaneously low latency, high mission success, and low communication overhead relative to the worst-case baseline.}

The 6G-LLM architecture demonstrates superior scalability, maintaining 80.7\% efficiency at 30 TADVNs compared to 84.5\% at 5 TADVNs. Conversely, the 5G-Traditional baseline remains ineffective, dropping to 17.0\%. Intermediate results reveal critical bottlenecks: 5G-LLM efficiency falls by 10.9 points as network constraints offset AI gains, while 6G-Traditional plateaus near 36\%, proving that hardware upgrades alone are insufficient. These results confirm that only the integrated 6G-LLM framework satisfies the rigorous reliability and scalability demands of next-generation tactical operations.

\begin{figure*}[t]
\centering
\begin{subfigure}[t]{0.48\textwidth}
    \centering
    \includegraphics[width=\linewidth]{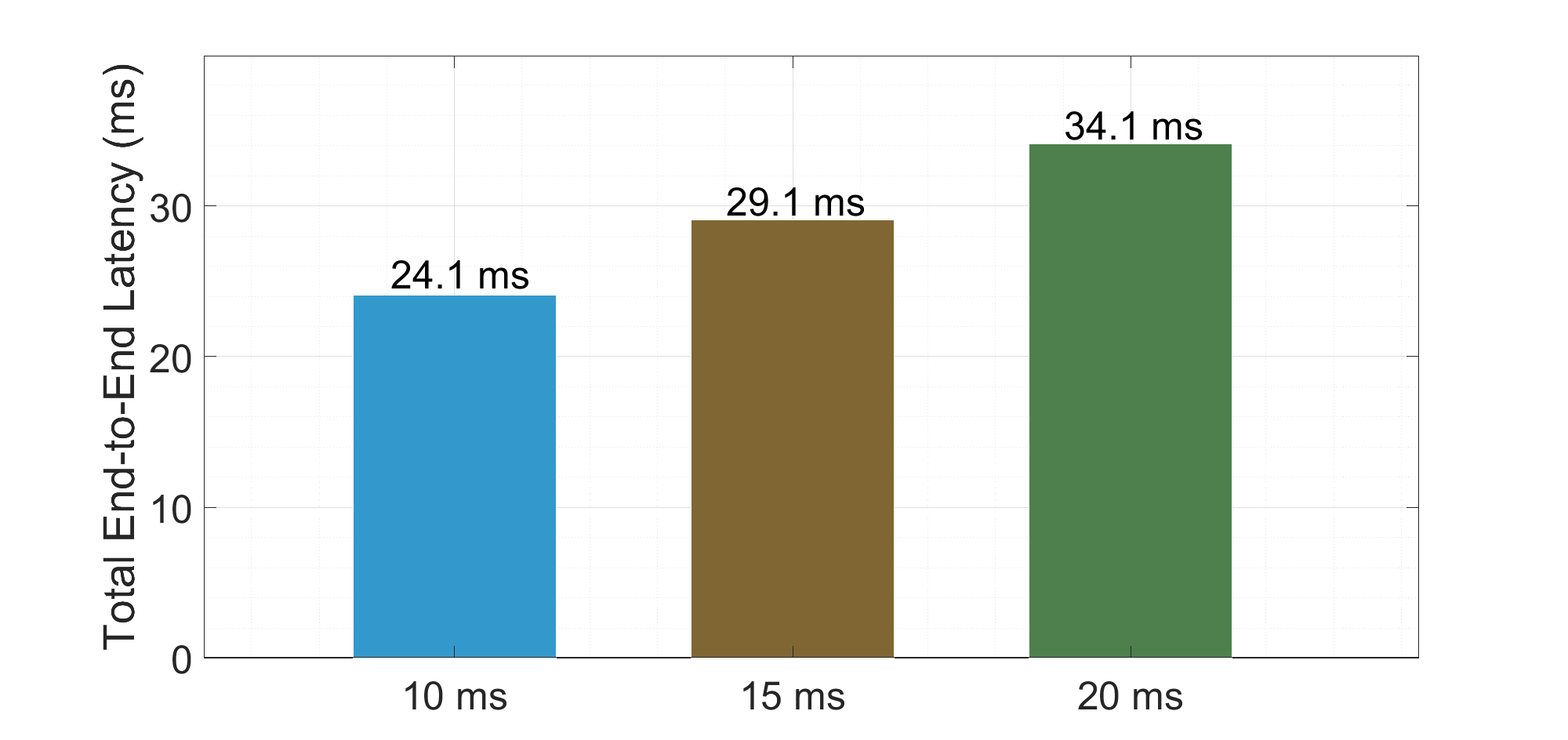}
    \caption{End-to-end latency sensitivity to LLM inference time under edge-assisted deployment.}
    \label{fig:llm_component}
\end{subfigure}
\hfill
\begin{subfigure}[t]{0.48\textwidth}
    \centering
    \includegraphics[width=\linewidth]{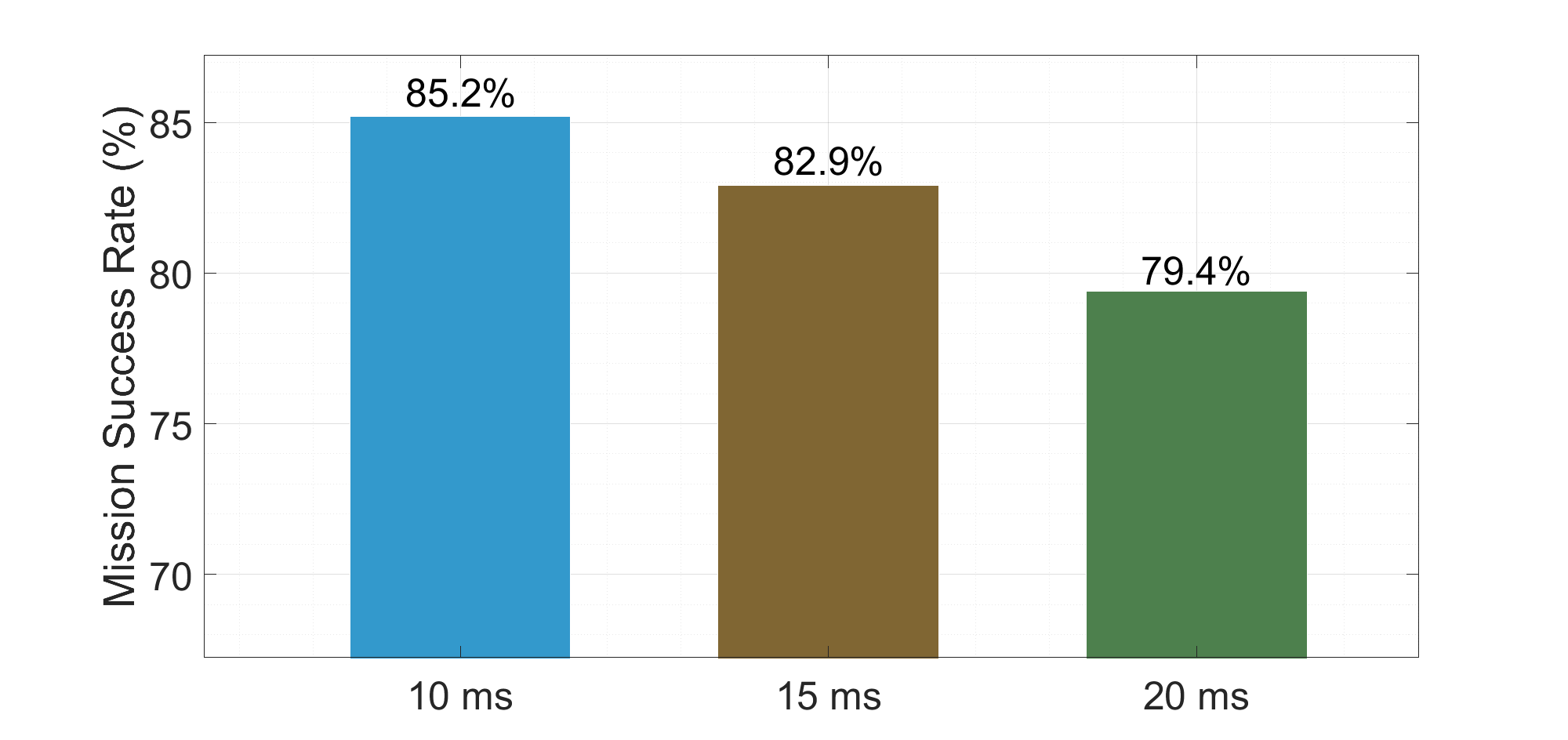}
    \caption{Mission success rate sensitivity to LLM inference time (30 TADVNs).}
    \label{fig:processing_comparison}
\end{subfigure}

\caption{Component-level evaluation of LLM processing and architectural delay contribution under edge-assisted deployment.}
\label{fig:llm_component_analysis}

\end{figure*}

{\color{black} 
\subsection{LLM Inference Delay Sensitivity Analysis}
To complement the system-level evaluation, we analyze how variations in LLM inference time influence overall coordination latency under edge-assisted deployment. The distilled 7B-parameter decoder-only transformer is executed with inference-time optimization consistent with the assumptions described in Section II.

Under edge GPU acceleration, the nominal LLM inference delay is approximately 15~ms. Internal profiling of this baseline configuration indicates that the total inference time consists of three stages: (i) input preprocessing and tokenization ($\approx$2~ms), (ii) transformer forward-pass computation ($\approx$10~ms), and (iii) semantic decoding and structured payload generation ($\approx$3~ms), with the transformer forward pass representing the dominant computational component. In the considered tactical scenarios, command inputs are short (typically $<$128 tokens), which bounds practical inference latency and prevents quadratic attention-scaling effects from dominating execution time.

Figure~\ref{fig:llm_component_analysis}(a) illustrates the sensitivity of total end-to-end latency to LLM inference time. As inference delay varies from 10~ms to 20~ms, overall coordination latency increases linearly, reflecting the additive architectural contribution of semantic reasoning to the delay budget. Even at 20~ms inference time, total latency remains well below the 100~ms operational threshold defined for time-critical tactical missions. These results indicate that LLM processing introduces a stable and predictable latency component within the proposed 6G+LLM architecture.

\subsection{Mission Success Robustness Under LLM Delay Variation}

We next evaluate how bounded variations in LLM inference delay affect large-scale coordination stability. Figure~\ref{fig:llm_component_analysis}(b) reports mission success rate at 30 TADVNs under inference delays of 10~ms, 15~ms, and 20~ms. Mission success degrades modestly as inference time increases (from 85.2\% at 10~ms to 79.4\% at 20~ms), demonstrating that moderate increases in semantic processing delay do not induce catastrophic coordination failure. The nominal 15~ms configuration corresponds to an 82.9\% success rate, consistent with the fleet-scale results reported in Table~II.  This bounded performance degradation confirms that edge-assisted semantic reasoning remains operationally stable under realistic processing-delay variations. The results further indicate that coordination robustness arises from the synergistic integration of low-latency 6G transport and localized semantic abstraction, rather than from network or AI enhancements in isolation.
}


\subsection{Scalability Analysis}
Table \ref{tab:performance} compares scalability across fleet sizes. The 6G–LLM architecture exhibits sub-linear degradation: despite a 6$\times$ fleet increase, latency rises only 33.8\% and mission success remains above 82\%. This demonstrates stable, bounded performance within operational limits.

In contrast, 5G-Traditional shows super-linear collapse: latency exceeds 100 ms beyond 15 units, while success rates plummet 59.1 points to an ineffective 14\%. Scalability analysis identifies critical phase transitions: 5G-Traditional fails at 15–20 units due to coordination loss, and 5G-LLM declines beyond 20 units due to latency bottlenecks. Only the integrated 6G–LLM framework maintains reliable performance across the full scale, whereas conventional systems become unsuitable beyond 15-20 units.

s

\section{Conclusion}
     {This paper proposes and evaluates, within a simulation-based scope bounded by IMT-2030 target parameters and Monte Carlo trend analysis, a multilayer communication-centric framework for 6G-enabled TADVNs that integrates edge-assisted LLM reasoning and schema-constrained semantic communication. The quantitative results characterize architectural performance trends under idealized next-generation connectivity assumptions; they are not intended as protocol-level guarantees or empirically validated deployment 
benchmarks, but rather as evidence that the proposed integration of 6G transport and LLM-driven semantic abstraction produces qualitatively distinct scalability behavior compared to conventional and partial-upgrade baselines.} The architecture enables efficient high-volume sensor processing with reduced latency and overhead, supporting adaptive mission execution and resilient fleet coordination under dynamic battlefield conditions. Monte Carlo simulations demonstrate significant gains at a 30-vehicle scale: 75.2\% latency reduction (29.1 ms vs.\ 117.5 ms), 68.7-point improvement in mission success (82.9\% vs.\ 14.2\%), and 88.6\% overhead reduction (17.3 MB/s vs.\ 151.9 MB/s) compared to 5G-Traditional systems. ITU-compliant validation confirms statistical significance, with 6G--LLM maintaining sub-linear degradation as fleet size increases sixfold, whereas conventional architectures exhibit super-linear collapse. 
Future work will address adversarially robust semantic communication, federated learning for privacy-preserving coordination, and hybrid classical--quantum cryptography to ensure secure, reliable next-generation autonomous defense systems.

\bibliographystyle{IEEEtran}
\bibliography{ref}

@article{li2023advanced,
  title={Advanced scenario generation for calibration and verification of autonomous vehicles},
  author={Li, Xuan and Teng, Siyu and Liu, Bingzi and Dai, Xingyuan and Na, Xiaoxiang and Wang, Fei-Yue},
  journal={IEEE Transactions on Intelligent Vehicles},
  volume={8},
  number={5},
  pages={3211--3216},
  year={2023},
  publisher={IEEE}
}

@ARTICLE{10579547,
  author={Bariah, Lina and Debbah, Mérouane},
  journal={IEEE Wireless Communications}, 
  title={AI Embodiment Through 6G: Shaping the Future of AGI}, 
  year={2024},
  volume={31},
  number={5},
  pages={174-181},
  doi={10.1109/MWC.015.2300521}}

@ARTICLE{10643253,
  author={Javaid, Shumaila and Khan, Naveed and Alwarafy, Abdulmalik and Saeed, Nasir},
  journal={IEEE Wireless Communications}, 
  title={{AGI and LLM-D}riven Spectrum Intelligence in Future Wireless Networks}, 
  year={2025},
  volume={},
  number={},
  pages={1-10},
  doi={10.1109/MWC.2025.3600789}}

@ARTICLE{9782734,
  author={Anzalone, Luca and Barra, Paola and Barra, Silvio and Castiglione, Aniello and Nappi, Michele},
  journal={IEEE Transactions on Intelligent Transportation Systems}, 
  title={An End-to-End Curriculum Learning Approach for Autonomous Driving Scenarios}, 
  year={2022},
  volume={23},
  number={10},
  pages={19817-19826},
  doi={10.1109/TITS.2022.3160673}}

@ARTICLE{10529728,
  author={Hazra, Abhishek and Munusamy, Ambigavathi and Adhikari, Mainak and Awasthi, Lalit Kumar and P, Venu},
  journal={IEEE Communications Standards Magazine}, 
  title={6G-Enabled Ultra-Reliable Low Latency Communication for Industry 5.0: Challenges and Future Directions}, 
  year={2024},
  volume={8},
  number={2},
  pages={36-42},
  doi={10.1109/MCOMSTD.0004.2300029}}

@ARTICLE{10745245,
  author={Kalor, Anders E. and Durisi, Giuseppe and Coleri, Sinem and Parkvall, Stefan and Yu, Wei and Mueller, Andreas and Popovski, Petar},
  journal={Proceedings of the IEEE}, 
  title={Wireless 6G Connectivity for Massive Number of Devices and Critical Services}, 
  year={2024},
  volume={},
  number={},
  pages={1-23},
  doi={10.1109/JPROC.2024.3484529}}

@ARTICLE{10529954,
  author={Kakkavas, Grigorios and Diamanti, Maria and Karyotis, Vasileios and Nyarko, Kwame Nseboah and Gabriel, Matthias and Zafeiropoulos, Anastasios and Papavassiliou, Symeon and Moessner, Klaus},
  journal={IEEE Wireless Communications}, 
  title={5G Perspective Of Connected Autonomous Vehicles: Current Landscape and Challenges Toward 6G}, 
  year={2024},
  volume={31},
  number={4},
  pages={299-306},
  doi={10.1109/MWC.014.2300277}}

@ARTICLE{10529727,
  author={Kaushik, Aryan and Singh, Rohit and Dayarathna, Shalanika and Senanayake, Rajitha and Di Renzo, Marco and Dajer, Miguel and Ji, Hyoungju and Kim, Younsun and Sciancalepore, Vincenzo and Zappone, Alessio and Shin, Wonjae},
  journal={IEEE Communications Standards Magazine}, 
  title={Toward Integrated Sensing and Communications for 6G: Key Enabling Technologies, Standardization, and Challenges}, 
  year={2024},
  volume={8},
  number={2},
  pages={52-59},
  doi={10.1109/MCOMSTD.0007.2300043}}

@article{series2015imt,
  title={IMT Vision--Framework and overall objectives of the future development of IMT for 2020 and beyond},
  author={Series, M},
  journal={Recommendation ITU},
  volume={2083},
  number={0},
  pages={1--21},
  year={2015},
  publisher={Electronic Publication Geneva, Switzerland}
}

@article{series2017minimum,
  title={Minimum requirements related to technical performance for IMT-2020 radio interface (s)},
  author={Series, M},
  journal={Report},
  volume={2410},
  number={},
  pages={2410--017},
  year={2017},
  publisher={International Telecommunication Union Geneva}
}

@article{jiang2021road,
  title={The road towards 6G: A comprehensive survey},
  author={Jiang, Wei and Han, Bin and Habibi, Mohammad Asif and Schotten, Hans Dieter},
  journal={IEEE Open Journal of the Communications Society},
  volume={2},
  pages={334--366},
  year={2021},
  publisher={IEEE}
}

@article{barros2025solving,
  title={Solving AI Foundational Model Latency with Telco Infrastructure},
  author={Barros, Sebastian},
  journal={arXiv preprint arXiv:2504.03708},
  year={2025}
}

@inproceedings{trichias20246g,
  title={6G global landscape: A comparative analysis of 6G targets and technological trends},
  author={Trichias, Konstantinos and Kaloxylos, Alexandros and Willcock, Colin},
  booktitle={2024 Joint European Conference on Networks and Communications \& 6G Summit (EuCNC/6G Summit)},
  pages={1--6},
  year={2024},
  organization={IEEE}
}

@ARTICLE{11037762,
  author={Al Ridhawi, Ismaeel and Aloqaily, Moayad},
  journal={IEEE Network}, 
  title={AI-Driven Next-Generation Edge Computing: Current and Future Trends}, 
  year={2025},
  volume={39},
  number={6},
  pages={189-197},
  doi={10.1109/MNET.2025.3580540}}

@ARTICLE{10061645,
  author={Li, He and Ota, Kaoru and Dong, Mianxiong},
  journal={IEEE Wireless Communications}, 
  title={Learning IoV in 6G: Intelligent Edge Computing for Internet of Vehicles in 6G Wireless Communications}, 
  year={2023},
  volume={30},
  number={6},
  pages={96-101},
  doi={10.1109/MWC.017.2200089}}

\end{document}